\documentstyle[prl,aps,epsf]{revtex}

\def\be{\begin{equation}}
\def\ee{\end{equation}}
\def\bea{\begin{eqnarray}}
\def\eea{\end{eqnarray}}

\begin{document}

\title{Quantum Hall Dynamics on von Neumann Lattice}

\author{ K. ISHIKAWA, N. MAEDA, T. OCHIAI, and H. SUZUKI}

\address{Department of Physics,Graduate School of Science, 
Hokkaido University,060-0810 Sapporo,Japan \\
E-mail:name@particle.sci.hokudai.ac.jp}   


\maketitle
\begin{abstract}
Quantum Hall Dynamics is formulated on von 
Neumann lattice representation where electrons in Landau levels are 
defined on 
lattice sites and are treated systematically  like lattice fermions.
We give  a  proof of the integer
Hall effect, namely the Hall conductance is the 
winding number of the propagator in the momentum space and is 
quantized exactly as  integer multiple of $\frac{e^2}{h}$  
in quantum Hall regime of the system 
of interactions and disorders. This shows that a determination of 
 the fine structure constant from integer quantum Hall effect is in 
fact possible.
We  present also a unified   mean field theory 
of the fractional Hall effect  for the in-compressible quantum liquid 
states based on flux condensation and point out that the known 
Hofstadter butterfly 
spectrum of the tight binding model has a deep connection with the 
fractional 
Hall effect of the continuum electrons. 
Thus two of the most intriguing and important physical phenomena of 
recent years, the  integer Hall effect and fractional Hall effect 
are studied and are solved partly by von Neumann lattice 
representation.
\end{abstract}

\section{von Neumann lattice representation}
\subsection{Quantum Hall system }\label{subsec:prod}
Quantum Hall system is a system of two dimensional electrons in 
strong perpendicular magnetic field,and is realized in semiconductors.
The system shows intriguing physical phenomena such as the integer 
Hall effect and the fractional Hall effect. The Hall conductance agrees
with $N\times e^{2}/h$ or   $p/q\times e^{2}/h$ in finite parameter regions,
hence ,quantum Hall effect is used as a standard of resistance and precise 
determination of the fine structure constant provided the above relations
are exact.It is a theoretical issue to find out if the above relations are 
exact or not.

The fractional Hall effect shows that the ground state of 
many electron systems of certain fractional filling is unique and
is a quantum liquid of having large energy gap. The energy gap 
vanishes and the ground state has an enormous degeneracy in the absence of
interactions. To find the mechanism of forming this kind of incompressible
liquid by interactions is another issue for the theorists.

Von Neumann lattice is a spatial lattice which is defined from a complete
set of coherent states, i.e., eigenstates of annihilation operator. 
Von Neumann lattice representation preserves a spatial symmetry in
lattice form and is useful in studying the above problems. We have used it 
for studying the quantum Hall systems and for solving the quantum Hall 
dynamics.The proof of the integer Hall effect and others have been given.
\cite{r1,r2,r3,r10,r12} 

In quantum Hall system, it is convenient to decompose the electron
coordinates (x,y) into two sets of variables,guiding center variables
and relative coordinates variables. Guiding center variables (X,Y)
stand for the center coordinates of cyclotron motion and their commutation
relation becomes an imaginary number that is inversely proportional to the
external magnetic field. A minimum complete set of coherent states
in this space has discrete complex eigenvalues and is known as von Neumann
lattice representation. Relative coordinates ($\xi,\eta$) satisfy the 
equivalent commutation relation and the one body free Hamiltonian is 
proportional to the summation of squares of relative coordinates.
Hence the electron has a discrete eigenvalue of energy. Each energy level
is known as Landau level and its degeneracy is specified by the center 
variables. From commutation relation, the degeneracy per area is proportional 
to the magnetic field.  
 The coherent states are defined by
\begin{eqnarray}
&(X+iY)\vert\alpha_{mn}\rangle=z_{mn}\vert\alpha_{mn}\rangle,\\
&z_{mn}=(m\omega_x+n\omega_y)a,\nonumber
\end{eqnarray}
where $m,n$ are integers and $\omega_x,\omega_y$ are
complex numbers which satisfy
${\rm Im}[\omega_x^*\omega_y]=1$
and $z_{mn}$ is a point on the lattice site in the complex plane;
an area of the unit cell is $a^2$.
We call this lattice the magnetic von Neumann lattice.
With a spacing $ a=\sqrt{2\pi\hbar\over eB}$, the completeness of the set
$\{\vert\alpha_{mn}\rangle\}$ is ensured.
\cite{pb}
Fourier transformed states denoted by
\begin{equation}
\vert\alpha_{{\bf p}}\rangle=\sum_{m,n}e^{ip_xm+ip_yn}
\vert\alpha_{m,n}\rangle,
\end{equation}
are orthogonal, that is,
\begin{equation}
\langle\alpha_{{\bf p}}\vert\alpha_{{\bf p}'}\rangle=
\alpha({\bf p})\sum_{N}(2\pi)^2\delta({\bf p}-{\bf p}'
-2\pi{\bf N}).
\end{equation}
Here, ${\bf N}=(N_x,N_y)$ is a vector with  integer values 
and ${\bf p}=(p_x,p_y)$ is a momentum in the Brillouin zone (BZ), that is, 
$|p_x|,|p_y|\le\pi$. 
The function $\alpha({\bf p})$ is  calculated by using the Poisson 
resummation
formula as follows:
\begin{eqnarray}
\alpha({\bf p})&=&\beta({\bf p})^*\beta({\bf p}),
\label{alpha}\\
\beta({\bf p})&=&\left(2{\rm Im}\tau\right)^{1\over4}
e^{i{\tau\over4\pi}p_y^2}
\vartheta_1({p_x+\tau p_y\over2\pi}\vert\tau),
\label{beta}
\end{eqnarray}
where $\vartheta_1(z\vert\tau)$ is a theta function and
the moduli of the von Neumann lattice is defined by 
$\tau=-\omega_x/ \omega_y$.
The magnetic von Neumann lattice 
is parameterized by $\tau$.
To indicate the dependence on $\tau$, we sometimes use a notation 
such as 
$\beta({\bf p}\vert\tau)$. 
For $\tau=i$, the von Neumann lattice becomes a square lattice.
For $\tau=e^{i2\pi/3}$,
it becomes a triangular lattice.
Some properties of the above functions are presented in References
 .Whereas $\alpha({\bf p})$ satisfies the periodic boundary condition, 
$\beta({\bf p})$ obeys a nontrivial boundary condition
\begin{equation}
\beta({\bf p}+2\pi{\bf N})=e^{i\phi(p,N)}\beta({\bf p}),
\end{equation}
where
$\phi(p,N)=\pi(N_x+N_y)-N_y p_x$.
We can define the  orthogonal state  which is normalized with 
$\delta$-function as follows:
\begin{equation}
|\beta_{{\bf p}}\rangle={\vert\alpha_{{\bf p}}
\rangle\over\beta({\bf p})}.
\label{albe}
\end{equation}

It should be noted that the state $\vert\alpha_{{\bf 0}}\rangle$ is
a null state,  that is, $\sum_{m,n}\vert\alpha_{mn}\rangle=0$,
because $\beta(0)=0$.

The Hibert space of one-particle states is also spanned by the state 
$\vert f_l\otimes\beta_{{\bf p}}\rangle$.
We call the state $\vert f_l\otimes\beta_{{\bf p}}\rangle$ 
the momentum state of the von Neumann lattice.
The wave function of the state $\vert f_l\otimes\beta_{{\bf p}}\rangle$ 
in the spatial coordinate space is given in references.

The probability density 
 $|\langle {\bf x}|f_l\otimes \beta_{{\bf p}}\rangle |^2$  
is invariant under the translation
$(\tilde x,\tilde y)\to (\tilde x +aN_x,\tilde y +aN_y)$.
Thus, the momentum state is an extended state.

\subsection{Field Theoretical Formalism and Topological Formula 
of Hall Conductance}

In the preceding section 
we obtain  one-particle states based on 
 the von Neumann lattice, that is, 
the coherent  state 
$|f_l\otimes\alpha_{mn}\rangle$ and the momentum state
$|f_l\otimes\beta_{{\bf p}}\rangle$ .Now  we develop the field 
theoretical formalism based on the momentum state . 
From now on, we denote $\vert f_l\otimes\beta_{{\bf p}}\rangle$ 
as $|l,{\bf p}\rangle$. 
We expand the electron field operator in the form
\begin{equation}
\psi({\bf x})=\int_{{\rm BZ}}{d^2p\over(2\pi)^2}\sum_{l=0}
^{\infty}b_l({\bf p})\langle {\bf x}|l,{\bf p}\rangle.
\label{field}
\end{equation}
$b_l({\bf p})$ satisfies the anti-commutation relation
\begin{equation}
\{b_{l}({\bf p}),b^\dagger_{l'}({\bf p}')\}=\delta_{l,l'}
\sum_N (2\pi)^2\delta({\bf p}-{\bf p}'-2\pi{\bf N})e^{
i\phi(p',N)},
\end{equation}
and the same boundary condition as $\beta ({\bf p})$.
$b^\dagger_l$ and $b_l$ are creation and annihilation operators
which operate on the many-body states.
The free Hamiltonian is given by 
\begin{equation}
{\cal H}_0=\int d^2 x \psi^\dagger ({\bf x})\hat{H}_0 \psi({\bf x})
=\sum_{l}\int_{{\rm BZ}}{d^2 p\over (2\pi)^2}E_l
 b^\dagger_{l}({\bf p})b_{l}({\bf p}).
\end{equation}
The density and current operators  in the momentum space 
$j_{\mu}=(\rho,{\bf j})$  become
\begin{eqnarray}
j_{\mu}({\bf k})&=&\int_{{\rm BZ}}{d^2 p\over (2\pi)^2}
\sum_{l,l'}b^\dagger_{l}({\bf p})b_{l'}({\bf p}-a\hat{\bf k})
\nonumber \\
&&\langle f_{l}\vert {1\over 2}\{v_\mu ,e^{ik\cdot\xi}\}
\vert f_{l'}\rangle
e^{-{i\over4\pi}a\hat k_x(2p-a\hat k)_y},
\label{jmu}
\end{eqnarray}
Here, $v^\mu=(1,-\omega_c\eta,\omega_c\xi)$,
and  $\hat{k_i}=W_{ij}k_j$.
The explicit form of $\langle f_{l}\vert
e^{ik\cdot\xi}
\vert f_{l'}\rangle$ and $W_{ij}$ are given in references.

The free Hamiltonian ${\cal H}_0$ is diagonal in the above basis.
However, the density operator is not diagonal with respect to
the Landau level index.
This basis, which we call the energy basis, is convenient to
describe the energy spectrum of the system.
In another basis,
${\cal H}_0$ is not diagonal and the density operator is diagonal.
This basis, which we call  the current basis, is convenient to describe
the Ward-Takahashi identity and the 
topological formula  of  Hall conductance.
There is no basis in which both the Hamiltonian and the density 
are diagonal. This is one of  peculiar features in a
magnetic field.

The current basis is constructed as follows.
Using a unitary operator, we can diagonalize
the density operator in the Landau level indices.
We define the unitary operator
\begin{equation}
U^\dagger_{ll'}({\bf p})=\langle f_l\vert
e^{i{\tilde p}\cdot\xi/a-{i\over4\pi}p_xp_y}
\vert f_{l'}\rangle.
\end{equation}
By introducing a unitary transformed operator
$\tilde b_l({\bf p})=\sum_{l'}U_{ll'}({\bf p})b_l({\bf p})$,
the density operator is written in the diagonal form and
the current operator  becomes a simple form:
\begin{eqnarray}
\rho({\bf k})&=&\int_{{\rm BZ}}{d^2 p\over (2\pi)^2}\sum_
{l}\tilde b^\dagger_{l}({\bf p})\tilde b_{l}({\bf p}-a\hat{\bf k}).
\end{eqnarray}
$\tilde b_l$ and $\tilde b^\dagger_l$
satisfy the anti-commutation relation and  boundary condition.

Here we show the Ward-Takahashi identity and the topological formula 
of Hall conductance using the current basis. 
The one-particle irreducible vertex part  
$\tilde{\Gamma}^\mu$ is connected with  
the full propagator 
by the Ward-Takahashi identity. 
The identity has crucial roles in the
following derivation of the topological formula of Hall conductance.
The Ward-Takahashi identity in this case becomes 
\begin{equation}
\tilde{\Gamma}_{\mu}(p,p)
= {\partial \tilde{S}^{-1}(p) \over \partial p^{\mu}}.
\label{wt}
\end{equation}
In a theory without a magnetic field, Ward-Takahashi identity
gives a  relation that the state of the dispersion
$\epsilon(p)$ moves with the velocity
${\partial \epsilon(p) \over \partial p_i}$.
However in a magnetic field, we can not diagonalize both the current
and the energy simultaneously. Therefore, the Ward-Takahashi identity
 does not imply the relation.

In a gap region, it was proven that the Hall conductance
is obtained not only from the retarded  product of
the current correlation function (Kubo formula), but also
from the time-ordered  product of it.
From the time-ordered  product of the current correlation function,
the Hall conductance is given by the slope of $\pi^{\mu\nu}(q)$ at the 
origin and is written as 
\begin{equation}
\sigma_{xy}={e^2 \over 3!}\epsilon^{\mu\nu\rho}
\partial_{\rho}\pi_{\mu\nu}(q)|_{q=0}.
\label{hallc}
\end{equation}
If the derivative $\partial_\rho$ 
acts on the vertex with the external line attached, 
its contribution becomes zero owing to the epsilon tensor.
Therefore,  the case  that the derivative acts on the bare propagator 
is  survived.

By  the Ward-Takahashi identity,  
 $\sigma_{xy}$ is written as a topologically invariant expression of 
the full propagator: 
\begin{eqnarray}
\sigma_{xy}&=&{e^2 \over h}{1\over 24\pi^2}\int_{{\rm BZ}\times S^1}
d^3 p \epsilon_{\mu\nu\rho}
\\
&& {\rm tr} \left(
\partial_\mu \tilde{S}^{-1}(p)\tilde{S}(p)
\partial_\nu \tilde{S}^{-1}(p)\tilde{S}(p)
\partial_\rho \tilde{S}^{-1}(p)\tilde{S}(p)\right)
\nonumber
\label{topological}
\end{eqnarray}
Here, the trace is taken over the Landau level index and the $p_0$ 
integral is a contour integral on a closed path.

Thus, we denote $S^1$ as the integration range. 
The integral  
${1\over 24\pi^2}\int tr(d\tilde{S}^{-1}\tilde{S})^3$ 
gives a integer value
under general assumptions and in fact counts the number or  Landau bands 
bellow the Fermi energy. 
Thus, the Hall conductance is proved to
be a integer times $e^2/h$. The impurities and interactions dot not 
modify the value of the $\sigma_{xy}$ if the Fermi energy is located in the
gap region or in the localized state region.

\section{Flux state mean field theory}
\subsection{flux state on von Neumann lattice}
We propose a new mean field theory based on the flux state on 
von Neumann lattices in this section. 

The dynamical flux which is generated by interactions 
plays an important role in our mean field theory. 
Dynamics is described by a lattice Hamiltonian, which is due to 
the external magnetic field, and by the induced  magnetic flux due to 
interaction, although the original electrons are defined 
on the continuum space. Consequently,
our mean field Hamiltonian is close to the Hofstadter Hamiltonian, 
which is a tight-binding model with uniform constant flux. 
For this reason there are similarities between their solutions.

Due to the two scales of periodicity, 
the Hofstadter Hamiltonian exhibits interesting structure as is seen in Figure.
\cite{r11} 
The largest gap exists along a line 
$\Phi=\nu\Phi_0$ with a unit of flux $\Phi_0$. 
The ground state energy becomes minimum also with this flux. 
These facts may suggest that the Hofstadter problem has some connection 
with the fractional Hall effect. 
We pursue a mean field theory of the condensed flux states in the 
quantum Hall system and point out that the Hofstadter problem is 
actually connected with the fractional Hall effect. 

\begin{figure*}
\vspace{-5cm}
\centerline{
\epsfysize=20cm\epsffile{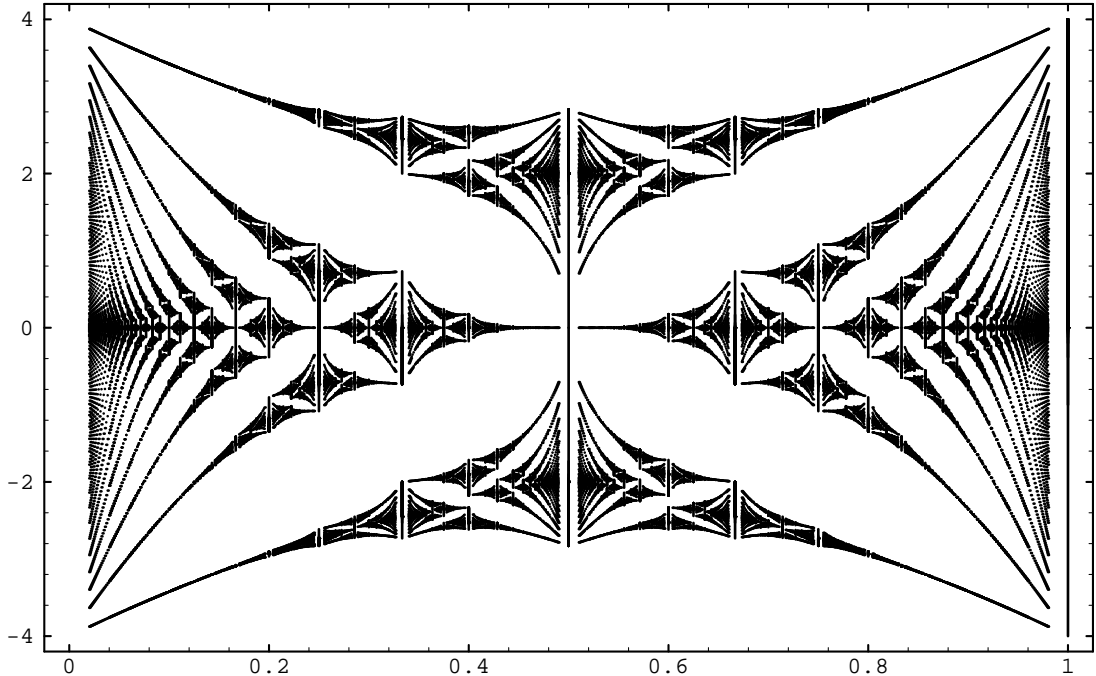}}
\vspace{-5cm}
\caption{The Hofstadter butterfly. Horizontal axis represents magnetic flux 
per plaquette. Vertical axis represents energy. }
\label{fig:radk}
\end{figure*}

$ $From the induced  magnetic field, new Landau levels are formed . 
If integer number of these Landau levels are filled completely, 
the integer quantum Hall effect occurs. 
The ground state has a large energy gap and is stable against 
perturbations, just as in the case of 
the ordinary integer quantum Hall effect. 
We identify them as fractional quantum Hall 
states. 

We postulate, in the quantum Hall system of the filling factor $\nu$, 
the induced  flux per plaquette and  magnetic field of 
the following magnitudes:
\begin{eqnarray}
&\Phi_{\rm ind}=\nu\Phi_0,\ \ \Phi_0=\Phi_{\rm external\ flux},
\nonumber\\
&B_{\rm ind}=\nu B_0,\ \ B_0=B_{\rm external\ magnetic\ field},
\end{eqnarray}
where $\nu$ is the filling factor measured with the external magnetic 
field. 
We obtain a self-consistent solution with this flux. 
Then the integer quantum Hall effect due to the induced magnetic field 
could occur just at the filling factor $\nu$, because the density 
satisfies the integer Hall effect condition, 
\begin{eqnarray}
&{eB_{\rm ind}\over 2\pi}N={eB_0\over 2\pi}\nu,\nonumber\\
&N=1.
\end{eqnarray}
The ground state has a large energy gap, generally. 

At the half-filling $\nu=1/2$, the half-flux $\Phi_0/2$ is induced. 
We first study the state of $\nu=1/2$, and next the states of  
$\nu=p/(2p\pm1)$. 
At $\Phi=\Phi_0/2$, the band structure is that of a massless Dirac 
field and has doubling symmetry. 
When an even number of Landau levels of the effective magnetic field, 
$B_{\rm ind}-B_0/2$, is filled, ground states have large energy gaps. 
This occurs if the condition of the density, 
\begin{eqnarray}
{e\over 2\pi}\vert\nu-{1\over2}\vert B_0\cdot 2p={eB_0\over2\pi}\nu,
\nonumber\\
\nu={p\over 2p\pm1}\ ;\ p,\ {\rm integer},
\end{eqnarray}
is satisfied. The factor of 2 on the left-hand side is due to the 
doubling of states and will be discussed later.
We study these states in detail based on the von Neumann lattice 
representation. 

The action and density operator 
show that there is an effective  magnetic field in the momentum space. 
The total flux in the momentum space is in fact a unit flux. 
In the thermodynamic limit, in which the density in space is finite, 
the density in momentum space is infinite. 
Consequently, it is possible to make this phase factor disappear 
using a singular gauge transformation in the momentum space with 
infinitesimally small coupling. 

We make a singular gauge transformation of the field in the momentum 
space and we have 
the commutation relation and the 
charge density.   

By a Chern-Simons gauge theory in momentum space, the gauge 
transformation is realized. 
Here, the coupling constant $\tilde e$ is infinitesimally 
small, and hence fluctuations of the Chern-Simons gauge field have 
a small effect, and we ignore them. 

The most important part in the action in the coordinate representation 
is obtained and
the corresponding Hamiltonian in the lowest Landau level space is given by
\begin{eqnarray}
&H=-{1\over 2}\sum v({\bf R}_2-{\bf R}_1) c_0^\dagger({\bf R}_1)
c_0({\bf R}_2) c_0^\dagger({\bf R}_2) c_0({\bf R}_1),
\nonumber\\
&v({\bf R})=
{\pi\over a}e^{-{\pi\over2}{\bf R}^2}I_0({\pi\over2}{\bf R}^2),
\end{eqnarray}
where $I_0$ is zero-th order modified Bessel function. 
We study a mean field solution of this Hamiltonian.

\noindent
(i) Half-filled case, $\nu=1/2$ 

At half-filling $\nu=1/2$, the system has a half flux, 
$\Phi=\Phi_0/2$. 
The system, then, is described equivalently with the two-component 
Dirac field by combining the field at even sites with that at 
odd sites. 
We obtained self-consistent solutions numerically. 
As is expected, the spectrum has two minima and two zeros corresponding 
to doubling of states. 

Around the minima, the energy eigenvalue  is approximated as,
\begin{equation}
E(p)=E_0+{({\bf p}-{\bf p}_0)^2\over 2m^*},\ \ 
{\bf p}_0=(0,0),(0,\pi/a),
\end{equation}.
The $m^*$ in Eq.(21) 
is the effective mass, 
and is  computed numerically:
\begin{equation}
m^*=0.225\sqrt{B\over B_0}m_e,\ B_0=20{\rm Tesla},\ \kappa=13,\ 
\gamma=0.914{e^2\over\kappa}.
\end{equation}

The $\nu=1/2$ mean field Hamiltonian is invariant under a kind of 
Parity, $P$, and 
anti-commutes with a  chiral transformation, $\alpha_5$. 
If the parity is not broken spontaneously, there is a degeneracy 
due to the parity doublet. 
The doubling of the states also appears at $\nu\neq1/2$ 
and plays an important role 
when we discuss the states away from $\nu=1/2$ in the next part. 
When an additional vector potential with the same gauge, 
$A_x=0,\ A_y=Bx$, is added, the Hamiltonian satisfies the  properties 
under the above transformations, and 
doubling due to the parity doublet also appears. Thus,
the factor of 2 is necessary in Eq.(19) and leads the principal 
series at $\nu=p/(2p\pm1)$ to have the maximum energy gap. 

\ 

\noindent
(ii) $\nu={p\over 2p\pm1}$

If the filling factor, $\nu$, is slightly away from 1/2, the 
total system can be regarded as a system with a small magnetic field 
of magnitude $(\nu-1/2)B_0$. 
The band structure may be slightly modified. 
It is worthwhile to start from the band of $\nu=1/2$ as a first 
approximation and to make iteration in order to obtain self-consistent 
solutions at arbitrary $\nu=p/(2p\pm1)$. 
We solve the mean field Hamiltonian 
\begin{eqnarray}
H_{\rm M}&=&\sum U^{({1\over 2}+\delta)}_0({\bf R}_1-{\bf R}_2)
e^{i\int({{\bf A}}^{({1\over 2})}+
\delta{\bf A})\cdot d{\bf x}}v({\bf R}_1-{\bf R}_2)
\nonumber  \\
&&c^\dagger({\bf R}_1)c({\bf R}_2),\nonumber \\ 
&&\langle c^\dagger({\bf R}_1)c({\bf R}_2)\rangle_{1/2+\delta}=
U^{({1\over 2}+\delta)}_0 e^{i\int({{\bf A}}^{({1\over 2})}
+\delta{\bf A})\cdot d{\bf x}}
\end{eqnarray}
under the self-consistency 
condition at $\nu=1/2+\delta$. 
Here we solve, instead, a Hamiltonian which has the 
phase of Eq.(23) but has the magnitude of the $\nu=1/2$ state. 

The integer quantum Hall state has an energy gap 
of the Landau levels due to $\delta{\bf A}$. 
This occurs when an integer number of Landau levels is filled 
completely. 
The Landau level structure is determined by the phase factor.
Magnitudes of the physical quantities  may be modified, 
nevertheless. 

$ U^{{1\over 2}}_0({\bf R}_1-{\bf R}_2)  $
was obtained in the 
previous part, 
and it is approximated with the effective mass formula. 

$ $From Eq.(19), 
$p$ Landau levels are completely filled, and the integer 
quantum Hall effect occurs at $\nu=p/(2p\pm1)$. 
The energy gap is given by the Landau level spacing 
\begin{equation}
\Delta E_{\rm gap}={e\delta B_{\rm eff}\over m^*}={eB_0\over m^*}
\vert\nu-{1\over2}\vert. 
\end{equation}

These equations were solved numerically, and the energy gaps and the 
widths of excited bands are obtained. 
Some bands are narrow and some bands are wide. 
Near $\nu=1/2$, the effective magnetic field approaches zero, and 
the Landau level wave functions have large spatial extension. 
The lattice structure becomes negligible, and the spectrum shows simple 
Landau levels of the continuum equation in these regions. 
Near $\nu=1/3$, the lattice structure is not negligible, and bands have 
finite widths. 
There are non-negligible corrections from those of continuum 
calculations. 

Due to the energy gap of the integer Hall effect caused by the induced 
dynamical magnetic field, the states at $\nu=p/(2p\pm1)$ are stable, 
and fluctuations are weak. 
Invariance under $P$, moreover, ensures these states to 
have uniform density. 
In systems with impurities, localized states with isolated discrete 
energies are generated by impurities and have energies in the gap regions. 
These states contribute to the density but do not contribute to 
the conductance. 
If the Fermi energy is in one of these gap regions, 
the Hall conductance $\sigma_{xy}$ is given by a topological formula, 
Eq.(16), and remains constant, at 
${e^2\over h}\cdot{p\over 2p\pm1}$. The fractional Hall effect is realized.
 
At a value of $\nu$ smaller than 1/3, 
the Hofstadter butterfly exhibits other kinds of 
structures. They may be connected with the Wigner crystall.

\subsection{Comparison with experiment}

In the previous section we presented our mean field theory based on 
flux condensation, where lattice structure generated by the external 
magnetic field and condensed flux due to interaction 
are important ingredients. 
Consequently, our mean field Hamiltonian becomes very similar to 
that of Hofstadter, which is known to show a large energy gap 
zone along the $\Phi=\nu\Phi_0$ line. 
The line $\Phi=\nu\Phi_0$ is special in the Hofstadter problem 
and hence in our mean field Hamiltonian. 
This explains why the experiments of the fractional quantum Hall effects 
show characteristic behavior at $\nu=p/(2p\pm1)$. 
The ground states at $\nu=p/(2p\pm1)$ have the lowest energy 
and the largest energy gap, and hence these states are stable. 
In this section we compare the energy gaps of the principal series 
in the lowest order, 
with experiments and with Laughlin variational wave function.

The effective mass $m^*$ of Eq.(22) 
was obtained from the curvature 
of the energy dispersion and should show a characteristic mass 
scale of the fractional Hall effect. 
The 
Landau level energy in the lowest approximation is calculated, and 
the gap energy is given in Eq.(24). 

These values are compared with the experimental values 
and 
with the composite fermion mean field values.  
Our mean field values are close 
to the experimental values. 
For example, at $\nu=1/3$, the composite fermion mean field theory 
gives $E_{\rm gap}$=380[K], which is a factor of 40 larger than the 
experimental value. 
Whereas, our effective mass formula gives $E_{\rm gap}$=24[K], and 
the other  approximation gives $E_{\rm gap}$=36[K], which are 
a factor of two or three larger than the experimental value and 
close to the value of the Laughlin wave function, $E_{\rm gap}$=26[K]. 
The agreement is not perfect, but should be regarded as good as the 
lowest mean field approximation. 
Near $\nu=1/3$, the bands have finite widths, and near $\nu=1/2$, the 
widths are infinitesimal. 
The dependence of the width upon the filling factor, $\nu$, 
and the entire structure of the bands   are 
characteristic features of the present mean field and 
should be tested experimentally. 

\section{Summary}
We formulated the quantum Hall effect, integer Hall effect and 
fractional Hall effect with the von Neumann lattice representation of 
two-dimensional electrons in a strong magnetic field. 
The von Neumann lattice is a subset of the coherent state. 
The overlap of states is expressed with a elliptic theta function. 
They allow for a systematic method of expressing 
the quantum Hall dynamics. 

A topological invariant expression of the Hall conductance was 
obtained in which compactness of the momentum space is 
ensured by the lattice of the coordinate space. 
Because the lattice has an origin in the external magnetic field, 
the topological character of the Hall conductance is ensured by 
the external magnetic field. 
The conductance is quantized exactly as $(e^2/h)\cdot N$ in the 
quantum Hall regime. 
The integer $N$ increases monotonically with the chemical potential. 

A new mean field theory of the fractional Hall effect that has 
dynamical flux condensation was proposed .
In our mean field theory, lattice structure is introduced from 
the von Neumann lattice and, flux is introduced dynamically. 
The mean field Hamiltonian becomes a kind of tight-binding model, and  
the rich structure of the tight-binding model is seen as 
characteristic features of the fractional Hall effect in our 
 mean field flux states of having  a liquid property  with an energy gap. 
These states satisfy the self-consistency condition of having 
the lowest energy and the largest energy gap. 
The physical quantities of our mean field theory are close to the 
experimental values in the lowest order at $\nu=p/(2p\pm1)$. 

\section*{Acknowledgements}
This work was partially supported by the special Grant-in-Aid 
for Promotion of Education and Science in Hokkaido University 
provided by the Ministry of Education, Science, Sports and Culture, 
the Grant-in-Aid for Scientific Research on Priority Area(Physics of CP 
violation) (Grant No.10140201), and the 
Grant-in-Aid for International Scientific Research (Joint Research
Grant No.10044043) from 
the Ministry of Education, Science, Sports and Culture, Japan.


\section*{References}
\begin{thebibliography}{99}
 
\bibitem{r1} 
N. Imai, K. Ishikawa, T. Matsuyama and I. Tanaka, Phys. Rev. {\bf B42} 
(1990), 10610.
\bibitem{r2}
K. Ishikawa, Prog. Theor. Phys. Supple. No.107 (1992), 167.
\bibitem{r3}
K. Ishikawa, N. Maeda and K. Tadaki, Phys. Rev. {\bf B51} (1995), 5048; 
{\bf B54} (1996), 17819.
K. Ishikawa et al., Phys. Lett. {\bf 210A} (1996), 321.
\bibitem{r10}
K. Ishikawa and N. Maeda, Prog. Theor. Phys. {\bf97} (1997),507 .
\bibitem{r11}
D.R.Hofstadter,Phys.Rev.{\bf B14},(1976),2239
\bibitem{r12}
K. Ishikawa, N. Maeda,T.Ochiai and H.Suzuki, 
Phys. Rev. {\bf B58} (1998), 1088;to appear in Physica E(1998);to appear in
Phys.Rev.B(1998).
\bibitem{pb} A. M. Perelomov, Teor. Mat. Fiz. {\bf 6}, 213 (1971);
V. Bargmann, P. Butera, L. Girardello, and J. R. Klauder,
Rep. Math. Phys. {\bf 2}, 221 (1971). 

\end{thebibliography}
\end{document}